\documentclass{revtex4}
\usepackage{color}
\usepackage{hyper ref}

\def\giorno{14/11/2016}

\def\a{\alpha}
\def\b{\beta}

\def\de{\delta}   
\def\eps{\varepsilon}
\def\vphi{\varphi}
\def\la{\lambda}

\def\s{\sigma}

\def\om{\omega}

\def\vphi{\varphi}

\def\R{{\bf R}}

\def\Om{\Omega}

\def\pa{\partial}

\def\o+{\oplus}

\def\<{\langle}
\def\>{\rangle}

\def\({\left(}
\def\){\right)}
\def\[{\left[}
\def\]{\right]}
\def\=#1{\bar #1}
\def\~#1{\widetilde #1}
\def\wt#1{\widetilde #1}
\def\.#1{\dot #1}
\def\^#1{\widehat #1}
\def\"#1{\ddot #1}

\def\eeq{\end{equation}}
\def\beq{\begin{equation}}

\def\beql#1{\begin{equation} \label{#1}}

\def\eqref#1{(\ref{#1})}

\def\EOR{ \hfill $\odot$ \medskip}

\def\EOP{ \hfill $\triangle$ \medskip}

\begin{document}

\title{Random Lie-point symmetries of stochastic differential equations}

\author{Giuseppe Gaeta\footnote{Member of GNFM-INdAM}}
\affiliation{Dipartimento di Matematica, Universit\`a degli Studi
di Milano, via Saldini 50, I-20133 Milano (Italy) \\ {\tt
giuseppe.gaeta@unimi.it} }
\author{Francesco Spadaro} \affiliation{EPFL-SB-MATHAA-CSFT, Batiment MA -
Station 8, CH-1015 Lausanne (Switzerland)
\\ {\tt francesco.spadaro@epfl.ch}}


\date{\giorno}


\begin{abstract}
\noindent
We study the invariance of stochastic differential equations under
random diffeomorphisms, and establish the determining equations
for random Lie-point symmetries of stochastic differential equations, both in
Ito and in Stratonovich form. We also discuss relations with
previous results in the literature.
\end{abstract}

\maketitle

\section{Introduction}
\label{sec:intro}

Symmetry analysis of differential equations is a powerful and by
now rather standard tool in the study of \emph{deterministic}
nonlinear problems \cite{Olver1,symmref,CGbook}; its use in the
context of \emph{stochastic} differential equations
\cite{sderef,Oksendal,LArnold,Stroock} is comparatively much less
developed (a partial exception being the versions of Noether
theorem \cite{Olver1,noeref} for stochastic variational problems
\cite{varstoref}).

Albeit some concrete results exists, in particular concerned with
\emph{strongly conserved quantities} related to symmetries
\cite{Mis1,Mis2,AlbFei} and to the \emph{linearization} problem
\cite{Meleshko}, a great deal of activity has been so far devoted
to discussing what the suitable definition of symmetry would be in
the case of stochastic differential equations (SDEs in the
following); here we refer e.g. to
\cite{Mis1,Mis2,AlbFei,Meleshko,WSM,symmstoref,Kozlov,GRQ1,GRQ2}.
See also \cite{GGPR} for a review.

As in the case of deterministic differential equations, these
works considered smooth vector fields in the space of independent
and dependent variables (also called the phase space or manifold);
in fact, dealing with smooth vector fields is at the hearth of the
Sophus Lie approach, in that it allows to deal with infinitesimal
transformations and hence with linearized problems.

On the other hand, when dealing with SDEs one is from the
beginning considering an object which is \emph{not} just a smooth
vector field; the evolution described by a SDE can be described in
terms of random diffeomorphisms, i.e. a diffeomorphism which
depends on a random process. Thus in this context it would be
quite natural to consider invariance under the same class of
transformations.

Actually this is exactly what has been done by L.Arnold and P.Imkeller
in their seminal work on \emph{normal forms} for SDEs
\cite{ArnImk} (see also the book by L.Arnold \cite{LArnold});
as well known, the theory of (Poincar\'e-Dulac in the case of general dynamical systems, or Birkhoff-Gustavsson for Hamiltonian ones) normal forms \cite{ArnGMDE,refNF1} is
intimately connected with symmetry properties \cite{CGbook,ArnGMDE,refNF2}, so that the success of their approach suggests that one can follow the same path in discussing general symmetry properties of SDEs outside the
perturbation approach.

The goal of the present paper is indeed to apply the
Arnold-Imkeller approach to the analysis of symmetries of SDEs. We
will see this can be done without difficulties, and explicit
\emph{determining equations} for the symmetry of a given SDE can
be obtained. The formulation of these is the main contribution of
our paper. We will also consider some concrete examples and determine symmetries for them, choosing equations which have physical relevance.

We will assume the reader has some familiarity with the basic
concepts in the theory of symmetry of (deterministic) differential
equations (see e.g. \cite{Olver1,symmref,CGbook}), and also with the basics of
stochastic differential equations (see e.g. \cite{sderef}); as the
former may be not so familiar to readers primarily interested in
SDEs, we will very briefly go over basic concepts for standard
(that is, deterministic) symmetries of SDEs.

We will first consider simple class of symmetries, in order to
focus on the main point of our contribution, and only later on
discuss the most general case. This will make the paper a little
longer than it would be going directly to the most general case,
but we trust it will help the reader -- not to say that simple
symmetries seem to be the most useful in applications.

The \emph{plan of the paper} is as follows. After briefly
introducing the class of maps to be considered, i.e. random
diffeomorphisms (Sect.\ref{sec:maps}), we will first discuss
symmetries of SDEs in Ito form (Sects.\ref{sec:ito} and
\ref{sec:general}) in increasing generality and examples of these
(Sect.\ref{sec:exaito}), passing then to discuss the case of
Stratonovich SDEs (Sect. \ref{sec:strato}) and examples of these
(Sect.\ref{sec:exastrato}). We then discuss the relation between
symmetries of an Ito equation and of its Stratonovich counterpart
(Sect.\ref{sec:correspondence}); we also discuss the (lack of)
simple algebraic structure of symmetry generators for a given Ito
equation and the (existence of) the same structure for a given
Stratonovich equation (Sect.\ref{sec:Lie}). Finally we draw our
brief Conclusions in Sect.\ref{sec:conclu}.
\bigskip

All the functions and other mathematical objects (manifolds,
vector fields) to be considered will be assumed -- unless
differently stated -- to be smooth; by this we will always mean
$\mathcal{C}^\infty$.

We will always use (unless differently stated) the Einstein
summation convention; we will usually denote partial derivatives
w.r.t. the $t$ and $x$ variables, and later on also w.r.t. the
$w^k$ variables, by the shorthand notation $$ \pa_t \ := \ (\pa /
\pa t) \ , \ \ \ \pa_i \ := \  (\pa / \pa x^i) \ ; \ \ \^\pa_k \
:= \ (\pa / \pa w^k) \ . $$

\section{Allowed Maps}
\label{sec:maps}

\subsection{Random diffeomorphisms} \label{sec:rd}

Arnold and Imkeller \cite{ArnImk} define a \emph{near-identity
random map} $h : \Om \times M \to M$, with $M$ a smooth manifold
and $\Om$ a probability space, as a measurable map such that: \par\noindent (i)
$h (\om , . ) \in \mathcal{C}^\infty (M)$; \par\noindent (ii) $h(\om, 0) = 0$;
\par\noindent (iii) $(Dh) (\om , 0) = id$.
\medskip

Property (i) means that we can consider this as a family of
diffeomorphisms (i.e., passing to generators, of vector fields) on
$M$, depending on elements $\om$ of the probability space $\Om$.
The dependence is rather arbitrary, i.e. no request of smoothness
is present.

We will also refer to the generator of such a map, with a slight
abuse of notation, as a \emph{random diffeomorphism}. Note that
random diffeomorphisms (as well as random maps) only act in $M$,
i.e. they do not act on the elements of $\Om$.

In our case, $M = \R \times M_0$, with $\R$
corresponding to the time coordinate, is the phase manifold for
the system, while $\Om$ will be the path space for the
$n$-dimensional Wiener process $W(t) = \{ w^1 (t) , ... , w^n (t)
\}$.

Moreover, as suggested by the notation above, we should consider
$M$ as a fiber bundle over $\R$ (the fibers being $M_0$) and $h$
should not act on $\R$.

In the end, introducing local coordinates $x^i$ on $M_0$, we want
to consider random diffeomorphisms generated by vector fields of
the form \beq X \ = \ \tau (x,t;w) \, \pa_t \ + \ \vphi^i (x,t;w)
\, \pa_i \ . \eeq

A \emph{time-preserving} random diffeomorphism will be
characterized by having $\tau = 0$, while the
\emph{fibration-preserving} ones (with reference to the fibration
$M \to \R$) will be characterized by $\tau = \tau
(t)$.  One should also mention that
special care is needed when considering time changes which depend
on $x$ (which is itself a stochastic process) and/or $w$, which
are \emph{random time changes} \cite{RTC}.

We will start by considering ``simple'' (i.e. time-preserving)
random symmetries in order to tackle the key problem in the
simplest setting; later on (see Section \ref{sec:general}) we will
consider the general case.

\medskip\noindent
{\bf Remark 1.} In the literature one considers also
transformations directly\footnote{As opposed to the ``indirect''
action due to the modification of the Wiener process induced by
the action on the time variable; see below.} acting on the Wiener
processes as well; this is related to so called ``W-symmetries''
\cite{GRQ2}. We will consider also this class of transformations,
in which case one considers diffeomorphisms (in the extended space
$(x,t;w)$) generated by vector fields of the form $ X = \tau
(x,t;w) \pa_t  +  \vphi^i (x,t;w) \pa_i + h^k (x,t;w)  \^\pa_k$. \EOR

\subsection{Maps acting on the time variable}
\label{sec:timechange}

If  we consider vector fields which act on the time variable as
well, we should take into account that the Wiener processes $w^k
(t)$ are affected by a change in $t$. In the simplest case, this
action on $t$ will be just a ``global'' reparametrization of time,
i.e. will not depend on the $x^i (t)$ and $w^k (t)$
variables.\footnote{Actually, in order to end up with an equation
possibly of the same type (not to say about it being exactly the
same equation as the original one) the transformed processes
$\wt{w}$ should be only a function of the (transformed) time
$\wt{t}$, i.e. $\wt{w} = \wt{w} (\wt{t})$; this amounts to
requiring the transformation of time does \emph{not} depend on the
space coordinates $x^i$. On the other hand, it could depend on the
Wiener processes themselves.}

This situation was discussed, in the context of symmetries for
SDEs, in \cite{GRQ1} (see Appendix A there); we give a short
account of this discussion here for the sake of completeness.

The probability that a Wiener process $w (t)$ undergoes a change
$d w = z$ in the time interval $\theta = d t$  has a density
$$ dp (z,\theta) \ = \ \[ 1/\sqrt{2 \pi \theta} \] \ e^{- z^2 / \theta } \ d z \ . $$
Under a near-identity map (we will assume $\tau' < 1/\eps$ for all
$t$) \beq \label{eq:tmap} t \ \to \ s \ = \ t \ + \ \eps \ \tau
(t) \ , \eeq we have $\theta  =  d t  =  [1/(1 + \eps \tau')] d
s$; thus the density $dp$ should now be expressed in terms of
$\^\theta = d s = (1 + \eps \tau' ) d t $.

Instead of going through computations, we note that if we consider
$\zeta = \sqrt{1 + \eps \tau'} z$ and the stochastic process \beq
\label{eq:wmap}  \^w (s) \ = \ \sqrt{1 + \eps \tau'} \ w(s) \ ,
\eeq the probability that $\^w (s)$ undergoes a change $\zeta = d
\^w$ in the time interval $\theta = d s $ has a density
$$ d \^p (\zeta, \theta) \ = \ \[ 1 / \sqrt{2 \pi \theta} \] \ e^{- \zeta^2 / \theta } \ d \zeta \ . $$

Thus we conclude that the map \eqref{eq:tmap} induces the map
\eqref{eq:wmap} on the standard Wiener process.

In the case of $\tau = \tau (x,t)$ extra care should be paid:
in general this would produce a random non-smooth map, and only
those expressed as integrals should be allowed \cite{RTC} (the
integration has a regularizing role); proceeding in a formal way
as we will do in the following has indeed in general a formal
value, and the actual well-posedness of the considered maps should
be verified in each case.

When the considered non-autonomous map is acceptable, with
$\tau = \tau (x,t)$  (see e.g. Theorem 8.20 in the book by
Oksendal \cite{Oksendal}) or even $\tau =\tau (x,t;w)$, one
proceeds in a similar way and obtains exactly the same result (see
Section \ref{sec:general}). This implies in particular that under
\eqref{eq:tmap}, \beq \label{eq:dwmap} d w^k \ \to \ d w^k \ + \
\eps \, \frac12 \, \( \frac{d \tau}{d t}  \) \ d w^k \ := \ d w^k
\ + \ \eps \ \de w^k \ . \eeq


\section{Ito equations; simple symmetries}
\label{sec:ito}

We will consider stochastic differential equations in Ito form,
i.e. \beql{eq:ito} d x^i \ = \ f^i (x,t) \, d t \ + \ \s^i_{\ k}
(x,t) \, d w^k \ . \eeq In the following it will be convenient to
use the notation \beq \label{eq:triangle} \triangle u \ := \
\sum_{k=1}^n \ \frac{\pa^2 u}{\pa w^k \pa w^k} \
 + \ \sum_{j,k=1}^{n}(\s\s^{T})^{jk} \frac{\pa^{2} u}{\pa
x^j \pa x^k} \ := \ \triangle_w u \ + \ \triangle_x u  \ . \eeq
For a function depending only on the $(x,t)$ variables -- as in
the case of deterministic symmetries -- the first term vanishes
identically.

\subsection{Deterministic symmetries}

We will start by considering \emph{simple symmetries}; in the
deterministic case these are generated by vector fields \beq
\label{eq:X0} X \ = \ \vphi^i (x,t) \, \pa_i \ , \eeq while when
we look for simple random symmetries we mean those generated by a
vector field \beq \label{eq:Y} Y \ = \ \vphi^i (x,t;w) \, \pa_i \
. \eeq

The \emph{determining equations for simple deterministic
symmetries} of Ito equations (that is, for $\tau = 0$ and $\vphi =
\vphi (x,t)$) were determined in \cite{GRQ1} (see also \cite{GRQ2}
for extensions), and turned out to be, in the present notation,
\beq \label{eq:symmsmoothito}
\begin{cases}
{\pa_t \vphi^i \ + \ f^j \, (\pa_j \vphi^i) \ - \ \vphi^j \,
(\pa_j f^i) \ = \  - \,  \frac12 \, \triangle(\vphi^i ) & , \cr
\s^j_{\ k} \, (\pa_j \vphi^i) \ - \ \vphi^j \, (\pa_j \s^i_{\ k} )
\ = \  0 \ . \cr} \end{cases} \eeq

\subsection{Random symmetries}

We will now consider the case of simple random symmetries, i.e. for vector fields of the form \eqref{eq:Y}; under this we have $ x^i \to x^i + \eps \vphi^i (x,w)$, and hence
\begin{eqnarray*}
d x^i & \to & d x^i \ + \ \eps \ d \vphi^i \\ & &  = \  d x^i \ + \ \eps \[ (\pa_j \vphi^i) \, d x^j \ + \ (\pa_t \vphi^i) \, d t \ + \ (\^\pa_k \vphi^i) \, d w^k \ + \ \frac12 \, ( \triangle \vphi^i ) \, d t \] \ ; \\
f^i (x,t) & \to & f^i(x,t) \ + \ \eps \ (\pa_j f^i) \, \vphi^j  \ , \\
\s^i_{\ k} (x,t) & \to & \s^i_{\ k} (x,t) \ + \ \eps \ (\pa_j \s^i_{\ k} ) \, \vphi^j \ . \end{eqnarray*}
Plugging these into \eqref{eq:ito}, the latter is mapped into a new Ito equation
\beq d x^i \ = \ [ f^i (x,t) \ + \ \eps \, (\de f)^i (x,t) ] \, d t \ + \  \s^i_{\ k} (x,t) \ + \ \eps \, (\de \s)^i_{\ k} (x,t) ] \, d w^k \ , \eeq where the variations are given by
\begin{eqnarray*}
(\de f)^i (x,t) & = &  [\vphi^j (\pa_j f^i) \ - \ f^j (\pa_j \vphi^i) \ - \ \frac12 \, (\triangle \vphi^i) \ - \ (\pa_t \vphi^i) ] \ , \\
(\de \s^i{\ k} ) (x,t) & = & [ \vphi^j (\pa_j \s^i_{\ k} ) \ - \ \s^j_{\ k} (\pa_j \vphi^i) \ - \ (\^\pa_k \vphi^i) ] \ . \end{eqnarray*}

Thus the equations remains invariant if and only if, for all $i$
and $k$, \beq \label{eq:itosymm}
\begin{cases}{
(\pa_t \vphi^i) \ + \ f^j (\pa_j \vphi^i) \ - \ \vphi^j (\pa_j
f^i)  \ = \  - \ \frac12 \, (\triangle \vphi^i) & \cr (\^\pa_k
\vphi^i) \ + \ \s^j_{\ k} (\pa_j \vphi^i) \ - \ \vphi^j (\pa_j
\s^i_{\ k} ) \ = \  0  &  . \cr}  \end{cases} \eeq
These are the
\emph{determining equations for simple random symmetries} -- of
the form \eqref{eq:Y} -- for the Ito equation \eqref{eq:ito}.

Note that introducing the vector fields \beq \label{eq:ZVF} X \ :=
\ \pa_t \ + \ f^j \, \pa_j \ , \ \ \^X \ := \ \^\pa_k \ + \
\s^j_{\ k} \, \pa_j \ ; \ \ Y_\vphi \ := \ \vphi^j \, \pa_j \ , \
\ Z_\vphi  \ := \ (\triangle \vphi^j) \, \pa_j \ ,  \eeq the
\eqref{eq:itosymm} are simply rewritten as \beq
\label{eq:itosymmshort} [X,Y_\vphi ] \ = \ -  \, \frac12 \, Z_\vphi \ ;
\ \ [ \^X , Y_\vphi ] \ = \ 0 \ . \eeq

\medskip\noindent
{\bf Remark 2.} The only difference w.r.t. the determining
equations for deterministic symmetries \eqref{eq:symmsmoothito} is
the presence of the $\pa_k \vphi^i$ term in the second equation;
but one should however recall that -- despite the formal analogy
-- the term $\triangle \vphi^i$ does now also include derivatives
w.r.t. the $w^k$ variables (i.e. the $\triangle_w \vphi^i$ term),
which are of course absent in \eqref{eq:symmsmoothito}, where
actually $\triangle \vphi^i = \triangle_x \vphi^i$. \EOR

\section{Ito equations; general random symmetries}
\label{sec:general}

So far we have considered (invariance of SDEs under) maps generated by vector fields of the special forms  \eqref{eq:X0} or \eqref{eq:Y}. We want now to remove this limitation, and consider general vector fields in the $(x,t;w)$ space, i.e.
\beq \label{eq:Ygen} Y \ = \ \tau (x,t;w) \, \pa_t \ + \ \vphi^i (x,t;w) \, \pa_i \ + \ h^k (x,t;w) \, \^\pa_k \ . \eeq
Here we started to use, as mentioned above, the shorthand notation
\beq \^\pa_k \ := \ \pa  / \pa w^k \ . \eeq
We also write $X = \tau \pa_t + \vphi^i \pa_i$ for the restriction of $Y$ to the $(x,t)$ space.

\medskip\noindent
{\bf Remark 3.} Note that in \eqref{eq:Ygen} we are considering
also the possibility of direct action on the $w^k$ variables
(apart from the action induced by a change in time), as in the
approach to W-symmetries \cite{GRQ2}. As already pointed out
there, the requirement that the transformed processes $\^w^k (t) =
w^k (t) + \eps h^k (x,t,w)$ are still Wiener processes, implies
that $\^w^k = M^k_{ \ell} w^\ell$ with $M$ an orthogonal matrix,
and hence that necessarily \beq h^k \ = \ B^k_{\ \ell} (x,t;w) \,
w^\ell \eeq with $B$ a (real) antisymmetric matrix; see
\cite{GRQ2} for details. This will be assumed from now on. (Note
moreover that if $B$ does not depend on $w$ then $\triangle (h^k)$
reduces to its ``deterministic'' part.) \EOR

\medskip\noindent
{\bf Remark 4.} On physical grounds one would be specially
interested in the case where the change of time does not depend on
either the realization of the stochastic processes $w^k (t)$ or on
the spatial coordinates $x^i$, i.e. on fiber-preserving maps.
These will be obtained from the general case by simply setting
$\tau = \tau (t)$. It should also be noted that, beside any
physical considerations, a (non trivially) space dependent time
change would provide a process which is not absolutely continuous
w.r.t. the original one - thus definitely not of interest in the
present context. See also the brief discussion in Sect.\ref{sec:timechange}. \EOR

\subsection{The general case}

The vector field \eqref{eq:Ygen} induces -- taking into account
also the discussion of the previous Section \ref{sec:timechange}
and in particular eq.\eqref{eq:dwmap} -- the infinitesimal map
\begin{eqnarray*} x^i &\to& x^i \ + \ \eps \ \vphi^i (x,t;w) \ , \\
t &\to& t \ + \ \eps \ \tau (x,t;w) \ , \\
w^k &\to& w^k \ + \ \eps \ h^k (x,t;w) \ + \ \eps \, \de w^k \ , \end{eqnarray*}

With this, the Ito equation \eqref{eq:ito} will read
\beq \label{eq:ni0} d x^i \ = \
\[ f^i (x,t)  + \eps (\de f)^i (x,t,w) \]  d t \, + \, \[
\s^i_k (x,t) + \eps (\de \s)^i_k (x,t,w) \] d w^k \ ; \eeq
we do of course aim at obtaining explicit expressions for $\de f$
and for $\de \s$.

Working, as always, at first order in $\eps$, we have
\begin{eqnarray}
f^i [x +  \eps \vphi ,
t  + \eps \tau  ] &=& f^i (x,t) \ + \ \eps \(
\tau \, \frac{\pa f^i}{\pa
t} \ + \ \vphi^j \, \frac{\pa f^i}{\pa x^j} \) \ := \ f^i (x,t) \,+ \ \eps \, X[ f^i (x,t)] \ , \nonumber \\
\s^i_k [x + \eps \vphi , t + \eps
\tau  ] &=& \s^i_k (x,t) \ + \ \eps \( \tau \,
\frac{\pa\s^i_k}{\pa t} \ + \ \vphi^j \, \frac{\pa \s^i_k}{\pa x^j} \) \ := \ \s^i_k (x,t) \ + \ \eps \ X [ \s^i_k (x,t) ] \ . \label{eq:fc1}
\end{eqnarray}

The differentials $d \vphi^i$, $d \tau$, $d h^k$ should be computed by
the Ito formula; for a generic function $F(x,t;w)$ we have, making use of \eqref{eq:ito},
\begin{eqnarray}
d F &=& (\pa_t F) d t \ + \ (\pa_j F) d x^j \ + \ (\^\pa_k F) d w^k \ + \ \frac12 (\triangle F) d t \nonumber \\
&=& (\pa_t F) d t \ + \ (\pa_j F) [f^j d t + \s^j_{\ k} d w^k] \ + \ (\^\pa_k F) d w^k \ + \ \frac12 (\triangle F) d t \nonumber \\
&=& \[ (\pa_t F) + f^j (\pa_j F) + \frac12 (\triangle F) \] \, d t \ + \ \[ (\^\pa_k F) + \s^j_k (\pa_j F) \] \, d w^k \nonumber  \\
&=& L[F] \, d t \ + \ Y_k (F) \, d w^k , \label{eq:dF}
\end{eqnarray} where we have defined the Misawa vector fields
$Y_\mu$ and the second order operator $L$ by \beq Y_0 := \pa_t +
f^j \pa_j \ , \ \ Y_k := \^\pa_k + \s^j_{\ k} \pa_j \ ; \ \ L :=
Y_0 + \frac12 \triangle \ . \eeq The expressions for $d \vphi^i$,
$d \tau$, $d h^k$ are immediately obtained specializing
\eqref{eq:dF}: \beq \label{eq:fc2} d \vphi^i \ = \ L[\vphi^i] \, d
t \ + \ Y_k (\vphi^i) \, d w^k \ , \ \ d \tau \ = \  L[\tau] \, d
t \ + \ Y_k (\tau) \, d w^k \ , \ \ d h^k \ = \ L[h^k] \, d t \ +
\ Y_k (h^k) \, d w^k \ .  \eeq

Using \eqref{eq:fc1} and \eqref{eq:fc2} we can rewrite \eqref{eq:ni0} in the form
\begin{eqnarray}
d x^i \ + \ \eps \, d \vphi^i &=& [ f^i \ + \ \eps X(f^i)] \, (d t
+ \eps d \tau ) \ + \ [\s^i_k \ + \ \eps X (\s^i_k ) ] \, (d w^k +
\eps \de w^k + \eps d h^k ) \ . \end{eqnarray} We like to write
this in the form \beq d x^i \ = \ f^i (x,t) \, d t \ + \ \s^i_k
(x,t) \, d w^k \ + \ \eps \, \de F^i \ ; \eeq here, setting $\de
w^k = \psi d w^k$ (with $\psi = (1/2) (\pa_t \tau)$, see
\eqref{eq:dwmap}), we have
\begin{eqnarray*}
\de F^i &=& - \, d \vphi^i \ + \ f^i \, d \tau \ + \ X(f^i) \, d t \ + \ \s^i_k \, d h^k \ + \ \psi \, \s^i_k \, d w^k \ + \ X (\s^i_k ) \, d w^k \\
&=& f^i \[ L(\tau ) d t + Y_k (\tau) d w^k \] \ - \ \[ L (\vphi^i ) + Y_k (\vphi^i ) \] \ + \ X(f^i) d t \\ & & \ + \ X(\s^i_k ) d w^k \ + \ \psi \s^i_k d w^k \ + \ \s^i_m \[ L (h^m ) d t + Y_k (h^m) d w^k \] \\
&=& \[ X(f^i) \ - \ L (\vphi^i) \ + \ f^i \, L(\tau ) \ + \ \s^i_k \, L(h^k) \] \ d t \\
& & \ + \ \[ X (\s^i_k ) \ - \ Y_k (\vphi^i ) \ + \ f^i \, Y_k (\tau) \ + \ \s^i_m  \, Y_k (h^m ) \] \ d w^k \ . \end{eqnarray*}

We thus conclude that the determining equation for (random)
symmetries of the Ito equation \eqref{eq:ito} are \beq
\label{eq:deteqsito}
\begin{cases} {X(f^i) \ - \ L (\vphi^i) \ + \
f^i \, L(\tau ) \ + \ \s^i_k \, L(h^k) \ = \ 0 & , \cr  X (\s^i_k
) \ - \ Y_k (\vphi^i ) \ + \ f^i \, Y_k (\tau) \ + \ \s^i_m  \,
Y_k (h^m ) \ = \ - \frac12 \, (\pa_t \tau) \ \s^i_k & . \cr}
\end{cases} \eeq

These can also be finally rewritten, using the explicit form of
$L$ and $\psi$, as \beq \begin{cases} {X(f^i) \ - \ Y_0 (\vphi^i)
\ + \ f^i \, Y_0 (\tau) ) \ + \ \s^i_k \, Y_0 (h^k ) \ = \ \frac12
\, \[ \triangle (\vphi^i) \ + \ f^i \, \triangle (\tau) \ + \
\s^i_k \, \triangle (h^k ) \] & , \cr X(\s^i_k) \ - \ Y_k
(\vphi^i) \ + \ f^i \,  Y_k (\tau) \ + \ \s^i_m \, Y_k (h^m ) \ =
\ - \, \frac12 \, (\pa_t \tau ) \, \s^i_k & . \cr} \end{cases}
\eeq
Several special cases are considered in the following.

\medskip\noindent
{\bf Remark 5.} This is a system of $n + n^2$ linear equations for
the  $2 n + 1$ unknown functions $\{ \tau , \vphi^1 , ... ,
\vphi^n; h^1 , ... h^n \}$; these reduce to $n$ or $n+1$ functions
if we consider simple symmetries or at least symmetries not acting
directly on the $w$ variables. Thus the system is over-determined
for all $n > 1$, and in general we will have no symmetries; even
in the case there are symmetries, the equations are not always
easy to deal with, despite being linear, due to the dimension. For
$n = 1$ the counting of equations and unknown functions would
suggest we always have symmetries, but the solutions could be only
local in some of the variables. \EOR

\medskip\noindent
{\bf Remark 6.} The solutions to the determining equations should
then be evaluated on the flow of the evolution equation (the Ito
SDE); this can lead some function to get less general, or even
trivial; see Example 1 below. \EOR

\medskip\noindent
{\bf Remark 7.} We focused on the definition of random symmetries
of a SDE and on the determining equations for these; on the other
hand, we have not considered how the symmetries can be used in the
study of the SDE. The first use of symmetries for SDEs should be
through the introduction of symmetry-adapted coordinates; (see
Remark 8 in this respect). A more structured approach, relating
simple symmetries to reduction pretty much as for deterministic
equations, has been developed by Kozlov \cite{Kozlov} in the case
of deterministic symmetries of SDEs; we postpone investigation of
the possibility to extend his results to the framework of random
symmetries to future work. \EOR

\subsection{Special cases}

It is interesting to consider some special (simpler) cases.

\medskip\noindent
{\bf (1)} In the case of deterministic simple (time preserving)
vector fields, i.e. $\vphi = \vphi (x,t)$, $\tau = h = 0$, the
equations \eqref{eq:deteqsito} reduce to the
\eqref{eq:symmsmoothito} seen above.

\medskip\noindent
{\bf (2)} Similarly, in the case of simple random symmetries, i.e.
$\vphi = \vphi (x,t;w)$, $\tau = h = 0$, we get the equations
\eqref{eq:itosymm} derived above.

\medskip\noindent
{\bf (3)} If we consider the case of deterministic
fiber-preserving symmetries, i.e.  $\vphi = \vphi (x,t)$, $\tau =
\tau (t)$, $h = 0$, the equations \eqref{eq:symmsmoothito} reduce
to \beq \begin{cases} { \pa_t \vphi^i \ - \ \pa_t (\tau f^i) \ + \
f^j \, (\pa_j \vphi^i) \ - \ \vphi^j \, (\pa_j f^i) \ = \ - \,
\frac12 \, \triangle \vphi^i & , \cr
 \tau \, (\pa_t \s^i_k) \ + \ \vphi^j \, (\pa_j \s^i_k) \ - \ \s^j_k \, \pa_j \vphi^i \ = \
 - \, \frac12 \, (\pa_t \tau) \, \s^i_k &  . \cr}
 \end{cases} \eeq
These equations coincide with those derived in \cite{GRQ1}, see
Theorem 2 there.

\medskip\noindent
{\bf (4)} When considering W-symmetries of SDEs \cite{GRQ2} one
considered vector fields with, in the present notation, $\vphi =
\vphi (x,t)$, $\tau = \tau (t)$, $h = h (t,w)$. In this case the
equations \eqref{eq:symmsmoothito} reduce to
  \beq  \begin{cases}
{ \pa_t \vphi^i \ - \ \pa_t (\tau f^i) \ + \ f^j \, (\pa_j
\vphi^i) \ - \ \vphi^j \, (\pa_j f^i) \ - \ \s^i_k \, (\pa_t h^k)
\ = \ \frac12 \, \s^i_k \, \triangle (h^k) \ - \, \frac12 \,
\triangle \vphi^i & , \cr
 \tau \, (\pa_t \s^i_k) \ + \ \vphi^j \, (\pa_j \s^i_k) \ - \ \s^j_k \,
 \pa_j \vphi^i \ + \ \s^i_{\ m} \, (\^\pa_k h^m) \ = \ - \, \frac12 \, (\pa_t \tau) \, \s^i_k&
 .\cr}
 \end{cases} \eeq
These equations were already obtained in \cite{GRQ2}, see the
Corollary to Proposition 1 there.

\medskip\noindent
{\bf (5)} Let us consider the general case with $\vphi = \vphi (x,t;w)$, $\tau = \tau (t,w)$, $h = 0$. The equations \eqref{eq:symmsmoothito} are in this case
  \beq  \begin{cases}
   { \pa_t \vphi^i \ + \ f^j \, (\pa_j \vphi^i) \ - \ \vphi^j \, (\pa_j f^i) \ + \ \tau \, (\pa_t f^i) \ - \
   f^i \, (\pa_t \tau)  \ = \  \frac12 \, \[  f^i \, \triangle (\tau) \, - \, \triangle (\vphi^i ) \]
   \ , \cr
    \^\pa_k \vphi^i \ + \ \s^j_k \, (\pa_j \vphi^i ) \ - \ \vphi^j \, (\pa_j \s^i_k) \ - \ \tau \, (\pa_t \s^i_k )
    \ - \ f^i \, (\^\pa_k \tau) \ = \ \frac12 \, (\pa_t \tau) \, \s^i_k & . \cr}
    \end{cases} \eeq

\medskip\noindent
{\bf (6)} As mentioned above (see Remark 4) we are specially interested in the case where $\tau = \tau (t)$ while $\vphi$ and $h$ are in general form (up to the restriction on $h$ discussed in Remark 3). In this case the only simplifications in \eqref{eq:symmsmoothito} are, of course, in the terms involving $\tau$, and amount to $Y_0 (\tau ) = (\pa_t \tau)$, $Y_k (\tau) = 0$, and $\triangle (\tau ) = 0$. Thus in this case the \eqref{eq:symmsmoothito} reduce to
  \beq  \begin{cases}
{ X(f^i) \ - \ Y_0 (\vphi^i) \ + \ f^i \, (\pa_t \tau) ) \ + \
\s^i_k \, Y_0 (h^k ) \ = \ \frac12 \, \[ \triangle (\vphi^i) \ + \
\s^i_k \, \triangle (h^k ) \]  & , \cr X(\s^i_k) \ - \ Y_k
(\vphi^i) \ + \ \s^i_m \, Y_k (h^m ) \ = \ - \, \frac12 \, (\pa_t
\tau ) \, \s^i_k & . \cr} \end{cases} \eeq

For $h=0$ (i.e. excluding W-symmetries) these further reduce to
  \beq  \begin{cases}
{ \pa_t \vphi^i \ + \ f^j \, \pa_i \vphi^i \ - \ \vphi^j \, \pa_j f^i \ - \ \tau \, \pa_t f^i \ - \ (\pa_t \tau) \, f^i \ + \ \frac12 \triangle \vphi^i \ = \ 0 & , \cr
\^\pa_k \vphi^i \ + \ \s^j_{\ k} \, \pa_j \vphi^i \ - \ \vphi^j \, \pa_j \s^i_{\ k} \ - \ \tau \, \pa_t \s^i_{\ k} \ - \ \frac12 (\pa_t \tau) \, \s^i_{\ k} \ = \ 0 & . \cr} \end{cases} \eeq

\medskip\noindent
{\bf (7)} Finally, in applications one is often faced with $n$-dimensional system, depending on $n$ Wiener processes,
$$ d x^i \ = \ (M^i_{\ j} \, x^j) \, d t \ + \ \s^i_{\ j} \ d w^j (t) \ , $$
with $M$ and $\s$ constant matrices. It is also frequent that $\s$ is diagonal.

In this case (for general, i.e. non necessarily diagonal, $\s$)
the determining equations for simple random symmetries read
\beq
\begin{cases}
{(\pa_t \vphi^i) \ + \ M^j_{\ q} \, x^q \ (\pa_j \vphi^i ) \ - \
M^i_{\ j} \, \vphi^j \ + \ \frac12 \, \triangle \vphi^i \ = \ 0 &
, \cr (\^\pa_k \vphi^i) \ + \ \s^j_{\ k} \, (\pa_j \vphi^i ) \ = \
0 & . \cr} \end{cases} \eeq We start by considering the second set
of equations; assuming moreover that $\s$ is diagonal, $\s =
\mathrm{diag} (\la_1 , ... , \la_n )$, these yield
$$ \vphi^i \ = \ \vphi^i (z^1,....,z^n;t) \ , $$
where we have defined $z^k := x^k - \la_k w^k$ (no sum on $k$).
For functions of this form we get immediately (using again the
\emph{ansatz} on $\s$) that $\triangle \vphi = 0$, and hence the
first set of determining equations read simply \beq \frac{\pa
\vphi^i}{\pa t} \ + \ \( M^j_{\ k} \, x^k \) \, \( \frac{\pa
\vphi^i}{\pa z^j} \) \ = \ M^i_{\ j} \, \vphi^j \ . \eeq This is
equivalent to
\begin{eqnarray}
\frac{\pa \vphi^i}{\pa t} \ = \ M^i_{\ j} \, \vphi^j \ \
\mathrm{and} \ \  \( M^T \)_p^{\ j} \, \( \frac{\pa \vphi^i}{\pa
z^j} \) &=& 0 \ .
\end{eqnarray} The first set of equations implies that
$$ \vphi^i (z^1,...,z^n ; t ) \ = \ e^{(t - t_0) M} \ \vphi^i (z^1,...,z^n ; t_0 ) \ , $$
while the second one states that $\nabla \vphi^i$ is in the kernel of $M^T$.

\section{Examples I: symmetries of Ito equations}
\label{sec:exaito}

\subsection{Simple random symmetries}

We start by considering simple random symmetries of Ito equations;
in this case the relevant determining equations are
\eqref{eq:itosymm}. We will consider examples which were already
studied -- for what concerns deterministic symmetries -- in
\cite{GRQ1}, so that comparison with results in the deterministic
case is immediate.

\medskip\noindent
{\bf Example 1.} We start by considering a rather trivial example, i.e. $n=1$ and \beq \label{eq:example1} d x \ = \ \s_0 \ d w (t)  \eeq with $\s_0 \not= 0$. In this case we just have a system of two equations for the single function $\vphi = \vphi (x,t;w)$, and \eqref{eq:itosymm} read
\begin{eqnarray*}
\pa_t \vphi &=& - (1/2) \ \triangle \vphi  \\
\pa_w \vphi  &=& - \, \s_0 \ (\pa_x \vphi) \ . \end{eqnarray*}
The solution to the second of these is
$ \vphi = F (z , t )$, where $F$ is an arbitrary (smooth) function of $z := x - \s_0 w$ and $t$.
Plugging this into the first equation, we get
$$ \pa_t F \ = \ - \, \s_0^2 \ \pa^2_z F \ . $$
This is an autonomous linear equation, and it is readily solved (e.g. by considering the Fourier transform of $F$), showing that there are nontrivial simple random symmetries. Note however these will grow exponentially fast in time.

It should also be noted that $dz = 0$ on solutions to our equation \eqref{eq:example1}, see Remark 6.

\medskip\noindent
{\bf Example 2.} We consider another one-dimensional example, i.e.
\beq \label{eq:example2} d x \ = \ d t \ + \ x \ d w \ ; \eeq this was considered in \cite{GRQ1},
where it was shown it admits no deterministic symmetries. The
\eqref{eq:itosymm} read now
\begin{eqnarray*}
& & \pa_t \vphi \ + \ \pa_x \vphi \ = \ - (1/2) \, \triangle \vphi \\
& & \vphi \ - \ x \, (\pa_x \vphi) \ - \ (\pa_w \vphi) \ = \ 0 \ ; \end{eqnarray*}
the second equation yields
$$ \vphi (x,t,w) \ = \ x \ \psi(z,t) \ ,  \ \ \ \ \ z := x \ e^{- w} \ ; $$
inserting this in the first equation -- and recalling that now the coefficients of different powers of $x$ must vanish separately, as $\psi = \psi (t,z)$, we get two equations,
$$\psi \ + \ z \, \psi_z \ = \ 0 \ , \ \ 2 \, \psi_t \ + \ 3 \, z \, \psi_z \ + \ 2 \, z^2 \, \psi_{zz} \ = \ 0 \ . $$
Solving these, we have $\psi (z,t) = (c_1/z) \exp[- t/2]$, with $c_1$ an arbitrary constant; and hence we conclude that the equation \eqref{eq:example2} admits a simple random symmetry:
$$ \vphi(x,t,w) \ = \ e^{w - t/2} \ . $$

\medskip\noindent
{\bf Example 3.} We pass to consider examples in dimension two; we will write the vector indices (in $x$, $w$, $\vphi$) as lower ones in order to avoid any misunderstanding. The first case we consider is a system related to work by Finkel \cite{Finkel}, i.e.
\begin{eqnarray}
d x_1 &=& (a_1 / x_1) \, d t \ + \ d w_1 \nonumber \\
d x_2 &=& a_2 \, d t \ + \ d w_2 \ ; \label{eq:example3} \end{eqnarray}
here $a_1,a_2$ are two non-zero real constants.

The first set of \eqref{eq:itosymm} reads in this case
\begin{eqnarray}
& & \frac{a_1}{x_1^2} \, \vphi_1 \ + \ \pa_t \vphi_1 \ + \ \frac{a_1}{x_1} \, \pa_1 \vphi_1 \ + \ a_2 \, \pa_2 \vphi_1 \ + \ \frac12 \ \triangle \vphi_1 \ = \ 0 \ ; \nonumber \\
& & \pa_t \vphi_2 \ + \ \frac{a_1}{x_1} \, \pa_1 \vphi_2 \ + \ a_2 \, \pa_2 \vphi_2 \ + \ \frac12 \ \triangle \vphi_2 \ = \ 0 \ , \label{eq:ex3} \end{eqnarray}
while the second set of determining equations \eqref{eq:itosymm} reads
\begin{eqnarray*}
\frac{\pa \vphi_1}{\pa w_1} \ + \ \frac{\pa\vphi_1}{\pa x_1} &=& 0
\ , \ \ \frac{\pa \vphi_1}{\pa w_2} \ + \ \frac{\pa\vphi_1}{\pa
x_2} \ = \ 0 \ , \ \ \frac{\pa \vphi_2}{\pa w_1} \ + \
\frac{\pa\vphi_2}{\pa x_1} \ = \ 0 \ , \ \ \frac{\pa \vphi_2}{\pa
w_2} \ + \ \frac{\pa\vphi_2}{\pa x_2} \ = \ 0 \ . \end{eqnarray*}
These of course imply that, setting $z_k := x_k - w_k$,
$$ \vphi_1 (x_1,x_2,t;w_1,w_2) \ = \ \eta_1 (t,z_1,z_2) \ , \ \ \
\vphi_2 (x_1,x_2,t;w_1,w_2) \ = \ \eta_2 (t,z_1,z_2) \ . $$
Plugging these into the equations \eqref{eq:ex3},  and again recalling that -- as $\eta_i = \eta_i (t,z_1,z_2)$ -- the coefficient of different powers of $x_1$ must vanish separately, the first of those equations enforces
$$ \eta_1 (t,z_1,z_2) \ = \ 0 \ , $$ while in the second we get $\pa \eta_2 / \pa z_1 = 0$ and the equation reads
$$ \frac{\pa \eta_2}{\pa t} \ + \ a_2 \, \frac{\pa \eta_2}{\pa z_2} \ + \ \frac{\pa^2 \eta_2}{\pa z_2^2} \ = \ 0 \ . $$
Again this autonomous linear equation is readily solved, showing that there are simple random symmetries.

\medskip\noindent
{\bf Example 4.} Finally we will consider another two-dimensional example,
which is an Ornstein-Uhlenbeck type process related to the Kramers equation:
\begin{eqnarray}
d x_1 &=& x_2 \, d t \nonumber \\
d x_2 &=& - K^2 \, x_2 \, d t \ + \ \sqrt{2 K^2} \ d w (t) \ ; \label{eq:example4}
\end{eqnarray}
note we have here a single Wiener process $w(t)$, and correspondingly we will look for solutions $\vphi^i = \vphi^i (x_1,x_2,t;w)$.

It was shown in \cite{GRQ1} that this system admits some deterministic
symmetries; in particular there are the symmetries
$$ X_0 \ = \ \pa_t \ , \ \ X_1 \ = \ \pa_1 \ , \ \ X_2 \ = \ e^{- K^2 t} \ \[ \pa_1 \ + \ K^2 \, \pa_2 \] \ . $$

As in the previous example, we will start from the second set of equations in \eqref{eq:itosymm}; for our system these read
\begin{eqnarray*}
(\pa \vphi_1 / \pa w) &=& 0 \ , \ \ (\pa \vphi_1 / \pa x_2) \ = \
0 \ , \ \ (\pa \vphi_2 / \pa w) \ = \  0 \ ,  \ \ (\pa \vphi_2 /
\pa x_2) \ = \  0 \ . \end{eqnarray*} These of course rule out any
possible dependence on $w$, i.e. show that there is no simple
random symmetry.

\subsection{General random symmetries}

{\bf Example 5.} We will consider again the equations of Example 2, i.e.
\beq \label{eq:example5} d x \ = \ d t \ + \ x \, d w \ ; \eeq
we have seen this does not admit any deterministic symmetry but it admits one simple random symmetry. We will now check this admits some more general random symmetry; in order to keep computations simple, we will restrict to the time-independent case $\tau=0$ and $\vphi_t = h_t = 0$.

In this case the equations \eqref{eq:deteqsito} read
\begin{eqnarray*}
 x h_x + x^2 h_{xx} - \vphi_x  - \frac12 \( \vphi_{ww} + x^2 \vphi_{xx} + x h_{ww} \) &=& 0 \\
 \vphi - \vphi_w - x \vphi_x + x h_w + x^2 h_x &=& 0 \ . \end{eqnarray*}

The second equation requires
$$ \vphi (x,w) \ = \ x \ \( h(x,w)  + \eta (z) \) \ , \ \ \ \ \ \ z := w - \log (|x|) \ ; $$
plugging this into the first one we get
$$ - \eta (z) \ + \ \eta' (z) \ + \ \frac12 \, \eta' (z) \ - \ x \ , \eta''(z) \ = \ h (x,w) \ - \ x^2 \, h_x (x,w) \ . $$
Solutions to these are provided by
$$ h (x,w) \ = \ e^{1/x} \, \b (w) \ + \ k \ , \ \ \eta (z) = - k \ , $$
with $k$ an arbitrary constant and $\b$ an arbitrary smooth function.

The random symmetries we obtained in this way are
\beql{eq:symmexample5} Y \ = \ \[ x \, e^{1/x} \, \b (w) \] \, \pa_x \ + \ \[ e^{1/x} \, \b (w) \ + \ k \] \, \pa_w \ . \eeq

\medskip\noindent
{\bf Example 6.} We consider the system
\begin{eqnarray}
d x_1 &=& [1 - (x_1^2 +x_2^2)] \, x_1 \, d t \ + \ d w_1 \nonumber \\
d x_2 &=& [1 - (x_1^2 +x_2^2)] \, x_2 \, d t \ + \ d w_2 \ ; \label{eq:example6} \end{eqnarray}
this is manifestly covariant under simultaneous rotations in the $(x_1,x_2)$ and the $(w_1,w_2)$ planes \cite{GRQ2}.

In order to simplify (slightly) the computations, we will look for symmetries which are time-preserving and time-independent; that is, we assume $\tau = 0$, $(\pa_t \vphi^i ) = 0 = (\pa_t h^k)$. The first set of \eqref{eq:deteqsito} provides now
\begin{eqnarray*}
\frac{\pa \vphi_1}{\pa w_1} \ + \ \frac{\pa \vphi_1}{\pa x_1} &=&
\frac{\pa h_1}{\pa w_1} \ + \ \frac{\pa h_1}{\pa x_1} \ , \ \
\frac{\pa \vphi_1}{\pa w_2} \ + \ \frac{\pa \vphi_1}{\pa x_2} \ = \ \frac{\pa h_1}{\pa w_2} \ + \ \frac{\pa h_1}{\pa x_2} \ ; \\
\frac{\pa \vphi_2}{\pa w_1} \ + \ \frac{\pa \vphi_2}{\pa x_1} &=&
\frac{\pa h_2}{\pa w_2} \ + \ \frac{\pa h_2}{\pa x_1} \ , \ \
\frac{\pa \vphi_2}{\pa w_2} \ + \ \frac{\pa \vphi_2}{\pa x_2} \ =
\  \frac{\pa h_2}{\pa w_2} \ + \ \frac{\pa h_2}{\pa x_2} \ .
\end{eqnarray*} Setting $z_k :=  x_k  -  w_k$, these give
\begin{eqnarray*}
h_1 (x_1,x_2,w_1,w_2) &=& \vphi_1 (x_1,x_2,w_1,w_2) \ + \ \rho_1 (z_1 , z_2 ) \\
h_2 (x_1,x_2,w_1,w_2) &=& \vphi_2 (x_1,x_2,w_1,w_2) \ + \ \rho_2 (z_1 , z_2 ) \ ,  \end{eqnarray*}
where the $\rho_i$ are arbitrary smooth functions of $(z_1,z_2)$.

Plugging these into the first set of \eqref{eq:deteqsito} we obtain two equations involving $\vphi^i$ and derivatives of the $\rho^i$,
\begin{eqnarray*}
(1 - 3 x_1^2 - x_2^2) \, \vphi_1 \ - \ 2 x_1 x_2 \, \vphi_2 \ + \ x_1 (1 - x_1^2 - x_2^2) \frac{\pa \rho_1}{\pa z_1} \ + \ x_2 (1 - x_1^2 - x_2^2) \, \frac{\pa \rho_1}{\pa z_2} \ + \ \frac{\pa^2 \rho_1}{\pa z_1^2} \ + \ \frac{\pa^2 \rho_1}{\pa z_2^2} &=& 0 \ ; \\
(1 - 3 x_1^2 - x_2^2) \, \vphi_2 \ - \ 2 x_1 x_2 \, \vphi_1 \ + \ x_1 (1 - x_1^2 - x_2^2) \frac{\pa \rho_2}{\pa z_1} \ + \ x_2 (1 - x_1^2 - x_2^2) \, \frac{\pa \rho_2}{\pa z_2} \ + \ \frac{\pa^2 \rho_2}{\pa z_1^2} \ + \ \frac{\pa^2 \rho_2}{\pa z_2^2} &=& 0 \ . \end{eqnarray*}
These equations can then be solved for the $\vphi^i$ in terms of the $\rho^i$, yielding some complicate expression we do not report.
This shows we have random symmetries in correspondence with arbitrary functions $\rho_i (z_1,z_2)$.

When these are linear,
$$ \rho_1 \ = \ r_{10} \ + \ r_{11} \, z_1 \ + \ r_{12} \, z_2 \ ; \ \
\rho_2 \ = \ r_{20} \ + \ r_{21} \, z_1 \ + \ r_{22} \, z_2 \ , $$
and writing
$ \chi :=  [-  1  +  3  (x_1^2 +x_2^2)]$,
the resulting random symmetries are identified by
\begin{eqnarray*}
\vphi_1 &=& (1/\chi) \ \[ x_1 (1 - x_1^2 - 3 x_2^2) r_{11} + x_2 (1 - x_1^2 - 3 x_2^2) r_{12} + 2 x_1^2 x_2 r_{21} + 2 x_1 x_2^2 r_{22} \] \\
\vphi_2 &=& (1/\chi) \ \[ 2 x_1^2 x_2 r_{11} + 2 x_1 x_2^2 r_{12} + x_1 (1 - 3 x_1^2 - x_2^2) r_{21} + x_2 (1 - 3 x_1^2 - x_2^2) r_{22} \] \\
h_1 &=& (1/\chi) \  \[ r_{10}   \chi + r_{11}   (w_1 + 2 x_1^3 - 3 w_1 (x_1^2 + x_2^2)) \right.\\ & & \left. + r_{12}   (w_2 + 2 x_1^2 x_2 - 3 w_2 (x_1^2 +x_2^2)) + 2   x_1^2 x_2   r_{21} + 2   x_1 x_2^2   r_{22} \] \\
h_2 &=& (1/\chi) \ \[ r_{20}   \chi + 2   x_1^2 x_2   r_{11} + 2   x_1 x_2^2   r_{12} + (w_1 + 2 x_1 x_2^2 - 3 w_1 (x_1^2 +x_2^2))   r_{21} \right.\\ & & \left.  + (w_2 + 2 x_2^3 - 3 w_2 (x_1^2 + x_2^2)) \]  \ . \end{eqnarray*}

With the choice
$$ r_{10} = 0 , \ r_{20} = 0 \ ; \ \ r_{11} = 0 , \ r_{12} = 1 , \ r_{21} = - 1 , \ r_{22} = 0 $$
we get just simultaneous rotations in the $(x_1,x_2)$ and $(w_1,w_2)$ planes \cite{GRQ2}.

\medskip\noindent
{\bf Remark 8.} It may be interesting, also in view of Remark 7,
to change coordinates as suggested by the symmetry; we will set
$x_1  =  \rho  \cos (\vartheta)$, $x_2 = \rho \sin (\vartheta )$;
and similarly $ w_1 = \chi \cos (\la )$, $w_2 = \chi \sin (\la )$.
With these coordinates, the equations \eqref{eq:example6} read
simply
\begin{eqnarray*}
d \rho &=& (1 - \rho^2) \, \rho \, d t \ + \ \cos (\la - \vartheta) \, d \chi \ - \ \chi \, \sin (\la - \vartheta) \, d \la \ , \\
d \vartheta &=& (1/\rho) \ [ \sin (\la - \vartheta) \, d \chi \ + \ \chi \, \cos (\la - \vartheta) \, d \la] \ . \end{eqnarray*}
The invariance under simultaneous rotations in the $(x_1,x_2)$ and $(w_1,w_2)$ planes (i.e. simultaneous shifts in $\vartheta$ and $\la$) is now completely explicit.  \EOR


\section{Symmetries of Stratonovich equations}
\label{sec:strato}

So far we have considered SDE in Ito form; as well known, in some framework it is convenient to consider instead SDE in Stratonovich form, \beq \label{eq:strato} d x^i \ =
\ b^i (x,t) \, d t \ + \ \s^i_{\ k} (x,t) \circ d w^k (t) \ . \eeq
In particular, these behave ``normally'' under change of coordinates (on the other hand, the Stratonovich integral is not a martingale and its rigorous meaning is not immediate).

Stratonovich equations were also considered by pioneers in the
symmetry analysis of differential equations
\cite{Mis1,Mis2,AlbFei}; we are only aware of works dealing with
\emph{deterministic} symmetries of Stratonovich equations, so we
believe a short discussion of their random symmetries is also of
interest; this is given in the present Section.

\subsection{Simple symmetries}
\label{sec:stratosimple}

\subsubsection{Simple deterministic symmetries}

We will first consider (also in order to familiarize with the
notation) the action of a \emph{deterministic} vector field
\eqref{eq:X0} on Stratonovich equations.

Under the action of $X$, the equation \eqref{eq:strato} is mapped
into \beq d x^i \ + \ \eps \, d \vphi^i \ = \ (b^i + \eps \vphi^j
\pa_j b^i ) \, d t \ + \ (\s^i_{\ k} + \eps \vphi^j \pa_j \s^i_{\
k} ) \circ d w^k \ ; \eeq taking into account \eqref{eq:strato}
and expanding the term $d \vphi$, we have that terms of first
order in $\eps$ cancel out if and only if \beq (\pa_t \vphi^i) \,
d t \ + \ (\pa_j \vphi^i) \, d x^j \ = \ (\vphi^j \pa_j b^i) \, d
t \ + \ (\vphi^j \pa_j \s^i_{\ k} ) \circ d w^k \ ; \eeq if now we
substitute for $d x$ according to \eqref{eq:strato}, this yields $$ (\pa_t
\vphi^i) \, d t \ + \ (\pa_j \vphi^i) \, (b^j d t \, + \, \s^j_{\
m} \circ d w^m) \ = \ (\vphi^j \pa_j b^i) \, d t \ + \ (\vphi^j
\pa_j \s^i_{\ k} ) \circ d w^k \ , $$ which is finally rewritten
as $$ \[ \pa_t \vphi^i \ + \ b^j (\pa_j \vphi^i ) \ - \ \vphi^j
(\pa_j b^i) \] \, d t \ + \ \[ \s^j_{\ k} (\pa_j \vphi^i) \ - \
\vphi^j (\pa_j \s^i_{\ k} ) \] \circ d w^k \ = \ 0 \ . $$

The vanishing of this (for all realizations of the Wiener
processes $w^k$) is possible if and only if the $(n+n^2)$
equations \beq \label{eq:deteqsmstr}
\begin{cases} {\pa_t \vphi^i \
+ \ b^j (\pa_j \vphi^i ) \ - \ \vphi^j (\pa_j b^i)  \ = \ 0 &
(i=1,...,n) \cr \s^j_{\ k} (\pa_j \vphi^i) \ - \ \vphi^j (\pa_j
\s^i_{\ k} ) \ = \  0 & (i,k=1,...,n) \cr} \end{cases} \eeq are
satisfied. These are the \emph{determining equations} for the
simple deterministic symmetry generators -- of the form
\eqref{eq:X0} -- of the Stratonovich SDE \eqref{eq:strato}.

\medskip\noindent
{\bf Remark 9.} We can introduce, as suggested by Misawa
\cite{Mis1}, the $(n+1)$ vector fields \beq Z_0 \ := \ \pa_t \ + \
b^i (x,t) \, \pa_i \ ; \ \ Z_k \ := \ \s^i_{\ k} (x,t) \, \pa_i \
. \eeq With this notation, the determining equations
\eqref{eq:deteqsmstr} read simply \beq [X,Z_\mu] \ = \ 0  \ \ \ \
(\mu = 0,1,...,n ) \ . \eeq

\subsubsection{Simple random symmetries}

We will now consider again the equation \eqref{eq:strato}, but now
discuss its variation under a vector field of the form \eqref{eq:Y};
we will go through the
same computation as in the previous subsection.

Under the action of $Y$, the equation \eqref{eq:strato} is mapped
into \beq d x^i \ + \ \eps \, d \vphi^i \ = \ (b^i + \eps \phi^j
\pa_j b^i ) \, d t \ + \ (\s^i_{\ k} + \eps \vphi^j \pa_j \s^i_{\
k} ) \circ d w^k \ ; \eeq taking into account \eqref{eq:strato}
and expanding the term $d \vphi$, we have that terms of first
order in $\eps$ cancel out if and only if \beq (\pa_t \vphi^i) \,
d t \ + \ (\pa_j \vphi^i) \, d x^j \ + \ (\^\pa_k \vphi^i) \circ d
w^k \ = \ (\vphi^j \pa_j b^i) \, d t \ + \ (\vphi^j \pa_j \s^i_{\
k} ) \circ d w^k \ ; \eeq the last term in the l.h.s. is the only
difference with respect to the computation in the deterministic
case. Considering $x$ on the solutions to \eqref{eq:strato}, we
get $$
 (\pa_t \vphi^i) \, d t \ + \ (\pa_j \vphi^i) \,
(b^j d t \, + \, \s^j_{\ m} \circ d w^m) \ + \ (\^\pa_k \vphi^i)
\circ d w^k \ = \ (\vphi^j \pa_j b^i) \, d t \ + \ (\vphi^j \pa_j
\s^i_{\ k} ) \circ d w^k \ , $$ which is also rewritten as $$
\[ \pa_t \vphi^i \ + \ b^j (\pa_j \vphi^i ) \ - \ \vphi^j (\pa_j
b^i)
\] \, d t
\ + \ \[ (\^\pa_k \vphi^i ) \ + \ \s^j_{\ k} (\pa_j \vphi^i) \ - \ \vphi^j
(\pa_j \s^i_{\ k} )
\] \circ d w^k \ = \ 0 \ ,  $$ and the determining equations for the
\emph{random} simple symmetry generators (of the form
\eqref{eq:Y}) of the Stratonovich SDE \eqref{eq:strato} are
therefore \beq \label{eq:deteqrndstr}
\begin{cases} {\pa_t
\vphi^i \ + \ b^j (\pa_j \vphi^i ) \ - \ \vphi^j (\pa_j b^i)  \ =
\ 0 & (i=1,...,n) \cr  \^\pa_k \vphi^i \ + \ \s^j_{\ k} (\pa_j
\vphi^i) \ - \ \vphi^j (\pa_j \s^i_{\ k} ) \ = \ 0 & (i,k=1,...,n)
\ . \cr}
\end{cases} \eeq

\medskip\noindent
{\bf Remark 10.} In order to express this in compact terms, it is
convenient to modify slightly the definition of the (Misawa)
vector fields associated with the SDE; we will now write
\begin{eqnarray}
Y_0 &=& \pa_t \ + \ b^i (x,t) \, \pa_i \ = \ Z_0 \ , \ \ Y_k \ = \
\^\pa_k \ + \ \s^i_{\ k} (x,t) \, \pa_i \ = \ \^\pa_k \ + \ Z_k \
. \label{eq:Yk} \end{eqnarray} Then the determining equations
\eqref{eq:deteqrndstr} read simply \beql{eq:symmYk} [ Y , Y_\mu ]
\ = \ 0  \ \ \ \ (\mu = 0,1,...,n ) \ . \eeq

\subsection{Symmetries acting on the time variable}
\label{sec:stratotime}

The computations presented in Section \ref{sec:stratosimple} above can
be extended to cover the case where the considered transformations
act on time as well; in this case the discussion of Section \ref{sec:timechange} should be taken into account.

\subsubsection{Deterministic symmetries}

In the simpler case, i.e. a smooth transformation not depending on
the random variables (deterministic symmetries), the role of $X$
in \eqref{eq:X0} will be taken by \beq \label{eq:Z} Z \ = \ \tau
(t) \, \pa_t \ + \ \vphi^i (x,t) \, \pa_i \ . \eeq Note that under
this we get $t \to s = t + \eps \tau (t)$.

As discussed above (see Section \ref{sec:timechange}), $w^k (t)$ is mapped into
$\wt{w}^k (t) = \sqrt{1 + \eps \tau_t}  w^k (t)$, and hence $d \wt{w}^k = [ 1 + \eps (\tau_t / 2) ] d w^k$.
Making use of this fact, and proceeding in the same way as above,
we get at first order in $\eps$ \begin{eqnarray*} (\pa_j \vphi^i) \, d x^j \ +
\ (\pa_t \vphi^i) \, d t &=& [ \vphi^j (\pa_j b^i) \, + \, \tau
(\pa_t b^i) \, + \, (\pa_t \tau) b^i ] \, d t \ + \ [ \vphi^j (\pa_j
\s^i_{\ k}) \, + \, \tau (\pa_t \s^i_{\ k} ) \, + \, (1/2) (\pa_t \tau)
\s^i_{\ k} ] \circ d w^k \ . \end{eqnarray*} Substituting now for $d x^j$
according to \eqref{eq:strato}, we get
\begin{eqnarray*}
& & \[ b^j (\pa_j \vphi^i) + (\pa_t \vphi^i) - \vphi^j (\pa_j b^i)
- \tau (\pa_t b^i) - (\pa_t \tau) b^i \] \ d t \ + \  \[ \s^j_{\ k} (\pa_j \vphi^i) - \vphi^j (\pa_j \s^i_{\ k}) -
\tau (\pa_t \s^i_{\ k} ) - (1/2) \tau' \s^i_{\ k} \] \circ d w^k \
= \ 0 \ ; \end{eqnarray*} the determining equations are therefore
\beq \label{eq:deteqSsim}
\begin{cases} {(\pa_t \vphi^i) + b^j
(\pa_j \vphi^i)  - \vphi^j (\pa_j b^i) - \tau (\pa_t b^i) - (\pa_t
\tau) b^i \ = \ 0 & , \cr \s^j_{\ k} (\pa_j \vphi^i) - \vphi^j
(\pa_j \s^i_{\ k}) - \tau (\pa_t \s^i_{\ k} ) - (1/2) (\pa_t \tau)
\s^i_{\ k} \ = \ 0 & . \cr}
\end{cases} \eeq

\subsubsection{Random symmetries}

When we consider random symmetries the computations are slightly
more complex. Proceeding in the same way as earlier on, we obtain
the determining equations in the form \beq
\label{eq:deteqSgen}
\begin{cases} {(\pa_t \vphi^i) + b^j (\pa_j
\vphi^i)  - \vphi^j (\pa_j b^i) - \tau (\pa_t b^i) - (\pa_t \tau )
b^i \ = \ 0 & , \cr \^\pa_k \vphi^i + \s^j_{\ k} (\pa_j \vphi^i) -
\vphi^j (\pa_j \s^i_{\ k}) - \tau (\pa_t \s^i_{\ k} ) - (1/2)
(\pa_t \tau) \s^i_{\ k} \ = \ 0 & . \cr}
\end{cases} \eeq

\medskip\noindent
{\bf Remark 11.} If we want to express this in terms of
commutation properties, we introduce the vector fields
\begin{eqnarray}
Z_0 &=& \pa_t \ + \ b^i (x,t;w) \, \pa_i \ , \ \ Z_k \ = \ \^\pa_k
\ + \ \s^i_{\ k} (x,t;w) \, \pa_i \ ;
\end{eqnarray}
then the determining equations are rewritten
as
\begin{eqnarray}
\[ Z_0 , Z \] &=& \tau_t \, (\pa_t + b^i \, \pa_i ) \nonumber \\
\[ Z_k , Z \] &=& (1/2) \tau_t \s^i_{\ k} \, \pa_i \end{eqnarray}

\section{Examples II: Stratonovich equations}
\label{sec:exastrato}

{\bf Example 7.} Let us consider the equation
$$ d x \ = \ - \, x \, dt \ + \ x \circ d w \ ; $$
in this case the Misawa vector fields are
$$ Y_0 \ = \ \pa_t \ - \ x \, \pa_x \ ; \ \ Y_1 \ = \ \pa_w \ + \ x \, \pa_x \ . $$
The requirement that $X := \vphi (x,t,w) \pa_x$ commutes with both $Y_0$ and $Y_1$ yields
$$ \vphi (x,t,w) \ = \ e^{- t} \ \eta (z) \ , \ \ \ z := (e^w / x ) \ . $$

\medskip\noindent
{\bf Example 8.} Let us consider the system
\begin{eqnarray*}
d x_1 &=& - \, x_2 \, dt \ + \ \a \, x_1 \circ d w_1 \\
d x_2 &=& - \, x_1 \, dt \ + \ \a \, x_2 \circ d w_2 \ . \end{eqnarray*}

The Misawa vector fields are now
$$ Y_0 \ = \ \pa_t \ - \ x_2 \, \pa_1 \ + \ x_1 \, \pa_2 \ ; \ \
Y_1 \ = \ \^\pa_1 \ + \ \a r \, \pa_1 \ , \ \ Y_2 \ = \ \^\pa_2 \ + \ \a r \, \pa_2 \ . $$
Requiring the vector field
$$ X \ = \ \vphi^1 (x_1,x_2,t,w_1,w_2) \ \pa_1 \ + \ \vphi^2 (x_1,x_2,t,w_1,w_2) \ \pa_2 $$ to commute with $Y_1$ and $Y_2$ enforces
$$ \vphi^1 \ = \ x_1 \, \eta^1 (z_1,z_2,t) \ , \ \ \vphi^2 \ = \ x_2 \, \eta^2 (z_1,z_2,t) \ , $$ where we have defined
$ z_k := [(a w_k - \log|x_k|)/a]$. Requiring now that $X$ also commutes with $Y_0$, we get that actually it must be $\eta^1 = \eta^2 = c$; thus in conclusion the only simple random symmetry of the system under consideration is
$$ X \ = \ \pa_1 \ + \ \pa_2 \ ; $$
this is actually, obviously, a simple \emph{deterministic}
symmetry.

\medskip\noindent
{\bf Example 9.} We consider again the equation
$$ d x \ = \ d t \ + \ x \, d w \ , $$ as in Example 2 above. The corresponding Stratonovich equation is
$$ d x \ = \ \[ 1 \, - \, \frac{x}{2} \] \, d t \ + \ x \circ d w \ ; $$ the determining equations \eqref{eq:deteqrndstr} for simple random symmetries of this Stratonovich equation read
\begin{eqnarray*}
\pa_t \vphi \ + \ [1 - (x/2)] \, (\pa_x \vphi) \ + \ (1/2) \, \vphi &=& 0 \\
\pa_w \vphi \ + \ x \, (\pa_x \vphi) \ - \ \vphi &=& 0 \ . \end{eqnarray*}
It is immediate to check these, or more precisely the first of these, do \emph{not} correspond to the equations obtained in Example 2.
But, this set of equations does admit as solution
$$ \vphi (x,t,w) \ = \ c_0 \ \exp[w - t/2] \ , $$
which is just  the same solution we found in Example 2.

\medskip\noindent
{\bf Example 10.} When dealing with symmetries of Stratonovich equations, it is customary to consider the system, first introduced by Misawa \cite{Mis1},
\begin{eqnarray*}
d x_1 &=& (x_3 - x_2) \, d t \ + \ (x_3 - x_2) \, \circ \, d w \\
d x_2 &=& (x_1 - x_3) \, d t \ + \ (x_1 - x_3) \, \circ \, d w \\
d x_3 &=& (x_2 - x_1) \, d t \ + \ (x_2 - x_1) \, \circ \, d w \ ;
\end{eqnarray*}
it is well known -- and immediately apparent -- that this admits the simple symmetry generated by
$$ X = (1/2) (x_1^2 + x_2^2 + x_3^2) \ (\pa_1 \, + \, \pa_2 \, + \, \pa_3 )  $$
(and many others, as discussed by Albeverio and Fei \cite{AlbFei}). Note that this involves only one Wiener process, which will induce a non-symmetric expression for the equivalent Ito system.

Using \eqref{eq:itostratequiv}, the equivalent system of Ito equations turns out to be
\begin{eqnarray*}
d x_1 &=& (1/2) \, (3 x_3 - x_2 - 2 x_1) \, d t \ + \ (x_3 - x_2) \, d w  \\
d x_2 &=& (x_1-x_3) \, d t \ + \ (x_1 - x_3) \, d w \\
d x_3 &=& (x_2 - x_1) \, d t \ + \ (x_2 - x_1) \, d w \ .
\end{eqnarray*}

It is immediate to check that the determining equations \eqref{eq:itosymm} are \emph{not} satisfied by $X$; more precisely, the second set of \eqref{eq:itosymm} are (of course) satisfied, while the first set is not: in fact, we get (for all $i=1,2,3$)
$$ \pa_t \vphi^i \ + \ f^j \, (\pa_j \vphi^i) \ - \ \vphi^j \, (\pa_j f^i) \ + \ \frac12 \, (\triangle \vphi^i) \ = \  F (x) \ , $$
where we have written $$ F(x) \ = \ 2 \, x_1^2 \ + \ 3 \, x_2^2 \ + \ 3 \, x_3^2 \ - \ \( \frac52 \, x_1 x_2 \ + \ 3 \, x_2 x_3 \ + \ \frac52 \, x_1 x_3 \) \ . $$

\section{Symmetries of Stratonovich vs. Ito equations}
\label{sec:correspondence}

As well known, there is a correspondence between stochastic
differential equations in Stratonovich and in Ito form. In
particular, the Stratonovich equation \eqref{eq:strato} and the
Ito equation \eqref{eq:ito} are equivalent if and only if the
coefficients $b$ and $f$ satisfy the relation
\beql{eq:itostratequiv} f^i (x,t) \ = \ b^i (x,t) \ + \ \frac12 \[
\frac{\pa}{\pa x^k} (\s^T)^i_{\ j} (x,t) \] \, \s^{kj} \ := \ b^i
(x,t) \ + \ \rho^i (x,t) \ . \eeq Note this involve implicitly the
metric (to raise the index in $\s$); as we work in $\R^n$
we do not need to worry about this. Moreover, for $\s$ (and hence also $\s^T$) a constant matrix, we get $\rho = 0$ i.e. $b^i = f^i$.

Note also that $\s$ is the same in \eqref{eq:strato} and in
\eqref{eq:ito}; thus \eqref{eq:itostratequiv} can be used in both
directions. In particular, we can immediately use it to rewrite
the determining equations for symmetries (of different types) of
the Stratonovich equation \eqref{eq:strato} in terms of the
coefficients in the equivalent Ito equation.

One would be tempted to study symmetries of an Ito equation by
studying the symmetries of the corresponding Stratonovich one.
This would be particularly attractive in view of the fact that the
determining equations \eqref{eq:deteqSgen} for random symmetries of
Stratonovich equations are substantially simpler than the
determining equations \eqref{eq:deteqsito} for symmetries of Ito
equations; the same holds at the level of determining equations
for simple random symmetries, as seen by comparing
\eqref{eq:deteqrndstr} and \eqref{eq:itosymm}.

Unfortunately, this way of proceeding would give incorrect
results (as also shown by Example 10 above); this is already clear in the case of simple random -- and actually even deterministic -- symmetries, so that we will just discuss this case

In fact, the determining equations \eqref{eq:deteqrndstr} for
random symmetries of \eqref{eq:strato} are immediately rewritten
in terms of the coefficients $f^i$ of the equivalent Ito equation
as ($i,k=1,...,n$) \beq \label{eq:itosymm0}
\begin{cases} {\pa_t
\vphi^i \ + \ [ f^j (\pa_j \vphi^i) - \vphi^j (\pa_j f^i)] \ - \ [
\rho^j (\pa_j \vphi^i) - \vphi^j (\pa_j \rho^i)] \ = \  0 & , \cr
\^\pa_k \vphi^i \ + \ [ \s^j_{\ k} \, (\pa_j \vphi^i) \ - \
\vphi^j \, (\pa_j \s^i_{\ k} ) ] \ = \ 0 & ; \cr }
\end{cases} \eeq
where $\rho^i (x,t)$ is defined in \eqref{eq:itostratequiv}.

Note that the equations \eqref{eq:itosymm0} can be expressed in
the compact form \eqref{eq:symmYk} of commutation with the vector
fields $Y_\mu$ defined in \eqref{eq:Yk}, except that now the same
vector field $Y_0$ should now better (but equivalently) be defined
as \beq Y_0 \ = \ \pa_t \ + \ [f^i (x,t) + \rho^i (x,t)] \, \pa_i
\ . \eeq The determining equations for deterministic symmetries of
\eqref{eq:ito} are also obtained in the same way from
\eqref{eq:deteqsmstr}, or directly from the above
\eqref{eq:itosymm0} by setting to zero the derivatives with
respect to the $w^k$ variables.

However, it is immediate to check that the equations
\eqref{eq:itosymm0} do \emph{not} coincide with the correct
equations \eqref{eq:itosymm}. The difference is due to \beq
\delta^i \ := \ \vphi^j \, (\pa_j \rho^i) \ - \ \rho^j \, (\pa_j
\vphi^i) \ - \ \frac12 \,  \triangle \vphi^i \ \not= \ 0 \ . \eeq
Note that this inequality generally holds (for $\s_x \not= 0$)
even in one dimension, and even for \emph{deterministic} vector
fields (i.e. for $\^\pa_k \vphi^i \equiv 0$).

In fact, in the one-dimensional deterministic case we get \beq
\delta \ = \ \frac12 \ \[ \frac{\pa}{\pa x} \, \( \vphi \, \s \,
\s_x \ - \ \phi \, \s^2 \) \] \ = \ \frac12 \ \[ \vphi^2 \ \(
\frac{\pa^2}{\pa x^2} \, \frac{\s^2}{\vphi} \)  \] \ . \eeq

The non correspondence between the symmetries of an Ito equation
and of the corresponding Stratonovich equation might seem rather
surprising at first; however, first of all the notion of
correspondence between an Ito and the associated Stratonovich
equation is not so trivial, as discussed e.g. in the last chapter
of the book by Stroock \cite{Stroock} (see in particular
Sect.8.1.2 there), and second one should in any case not
expect identity of symmetries, but rather a \emph{correspondence}
between the two; thus the difference between the symmetries of
the two is not so strange.

On the other hand, an Ito equation and the associated Stratonovich
equation do carry the same statistical information. In view of the
discussion and results in \cite{GRQ1}, we would expect there is a
correspondence between symmetries of the Fokker-Planck equation
(symmetries of scalar Fokker-Planck equations were classified in
\cite{CicVit}, see also \cite{ShtSto}) which are also symmetries
of the Ito equation and symmetries of the equivalent Stratonovich
equation. This is indeed the case.

\medskip\noindent
{\bf Proposition.} {\it Given an Ito equation and the associated
Fokker-Planck equation, the symmetries of the latter which are
also symmetries of the Ito equation, are also symmetries of the
associated Stratonovich equation.}

\medskip\noindent
{\bf Proof.} In \cite{GRQ1} it was shown that symmetries of the Fokker-Planck equation
\beq \pa_{t} \, u \ + \ A^{ij} \ \pa_{ij}^{2} \, u \ + \ B^{i} \ \pa_{i} \, u \ + \ C \ u \ = \ 0 \eeq
with $A=-[(1/2) \sigma \sigma^{T} ]$, $B^i=f^i + 2 \pa_{j} A^{ij}$, $C= (\pa_{i}\cdot f^{i}) +\pa_{ij}^{2}A^{ij}$
have to satisfy the system \begin{eqnarray}
\sigma_{j}^{i}\Gamma^{k}_{s}\delta^{js}+\sigma^{k}j\Gamma^{i}_{s}\delta^{js} &=& 0 \nonumber \\
\Lambda^{i}+2\left[A^{ik}\pa_{k}\beta+A^{im}\pa_{mk}^{2}\xi^{k}\right] &=& 0 \\ \left[\pa_{t}+f^{i}\pa_{i}-A^{ik}\pa_{ik}^{2}\right]\left[\beta+\pa_{m}\xi^{m}\right] &=& 0 \nonumber
\end{eqnarray}
where
\begin{eqnarray}
\Gamma^{k}_{j} &=& \sigma_{j}^{m}\pa_{m}\xi^{k}-\xi^{m}\pa_{m}\sigma^{k}_{j}-\tau\pa_{t}\sigma^{k}_{j}-\frac{1}{2}\sigma^{k}_{j}\pa_{t}\tau
\nonumber \\
\Lambda^{i} &=& -\left[\pa_{t}(\xi^{i}-\tau f^{i})+\{f,\xi\}^{i}-A^{mk}\pa_{mk}^{2}\xi^{i}\right] \ .
\end{eqnarray}
The symmetry of the Fokker-Planck is also a symmetry of the Ito equation if and only if $\Gamma_{j}^{k}=0$ for all $j,k$, since the condition for a symmetry of the Ito equation are given by
$$
\Lambda^{i} \ = \ 0 \ , \ \ \ \Gamma_{j}^{k} \ = \ 0 \ .
$$
On the other hand, the symmetries of the Stratonovich equation are
given by \eqref{eq:itosymm0}, which can now be written as
\begin{eqnarray} \pa_{t}(\xi^{i}-\tau f^{i})+\pa_{t}(\tau \rho^{i}) \ + \{f,\xi\}^{i}- \{\rho,\xi\}^{i} &=& 0 \ , \nonumber \\
\Gamma^{k}_{j}&=& 0 \ ,
\end{eqnarray}
where $\rho^{i}=(1/2)(\pa/\pa x^k)[ (\s^T)^i_{\ j} \s^{kj}]$.

Thus it will suffice to show that
\begin{equation}
A^{mk} \ \pa_{mk}^{2} \xi^{i} \ + \ \pa_{t} \, (\tau \rho^{i}) \ - \ \{\rho,\xi\}^{i} \ = \  0 \ .  \end{equation}
An explicit computation shows that
\begin{eqnarray}
A^{mk}\pa_{mk}^{2}\xi^{i}+\pa_{t}(\tau \rho^{i}) \ - \{\rho,\xi\}^{i} &=&   \frac12\sum_{j}[\sigma_{j,},\Gamma_{j}]^{i} 
\ := \ \frac12\sum_{j}\sigma_{j}^{k}\pa_{k}\Gamma^{i}_{j}+\Gamma_{j}^{k}\pa_{k}\sigma_{i}^{j} \ ;
\end{eqnarray}
this completes the proof. \EOP

\section{Algebraic structure of symmetries}
\label{sec:Lie}

It is well known that the Lie-point symmetries (or more precisely
the Lie-point symmetry generators) for a given deterministic
differential equation form a Lie algebra
\cite{Olver1,symmref,CGbook}. One may wonder if the same holds in
the case of stochastic differential equation. The question was
answered in the negative by Wafo Soh and Mahomed \cite{WSM} for
(first order) Ito equations.

Here we want to discuss this point for Ito equations -- confirming
of course the result of \cite{WSM} -- and the analogous problem
for Stratonovich ones; we will be satisfied with a discussion in
the framework of simple symmetries (the negative result in the Ito
case will {\it a fortiori} hold for general symmetries).

\subsection{Ito equations}

We will consider vector fields $Y_\xi = \xi^j \pa_j$ and $Y_\eta =
\eta^j \pa_j$ which are symmetries for a given SDE; to these we
associate the vector fields $Z_\xi$ and $Z_\eta$ as in
\eqref{eq:ZVF}. By assumption we have \beq \label{eq:alg1} [ X ,
Y_\xi ] =  \frac12 Z_\xi , \ [ \^X , Y_\xi ] = 0 \ ; \ \ [ X
, Y_\eta ] = \frac12 Z_\eta \ , [ \^X , Y_\eta ] = 0 \ . \eeq
We now want to consider \beq \label{eq:alg2} Y_\vphi \ := \ \[
Y_\xi \ , \ Y_\eta \] \eeq and wonder if this is also a symmetry
for the same SDE.

It is immediate to check that $[\^X , Y_\vphi ] = 0$, just by Jacobi identity. As for the first of \eqref{eq:itosymmshort}, here Jacobi identity implies that
$ [ X \ , \ Y_\vphi ]  =  [ X , [ Y_\xi , Y_\eta ] ]  =  [ Y_\xi , [X , Y_\eta ] ] - [ Y_\eta , [X , Y_\xi ] ]$; using now \eqref{eq:alg1}, this reads
\beq \label{eq:alg3} \[ X \ , \ Y_\vphi \] \ = \ \frac12 \ \left\{ [ Y_\xi , Z_\eta ] \ - \ [ Y_\eta , Z_\xi ] \right\} \ . \eeq

In order to check if the first of \eqref{eq:itosymmshort} is
satisfied, we must express $Z_\vphi$ in terms of the $\{
Y_\xi,Y_\eta , Z_\vphi,Z_\eta \}$. Using \eqref{eq:alg2} and some
simple algebra, we get
\begin{eqnarray*} \triangle \vphi^i &=& (\triangle \xi^j ) \, \pa_j \eta^i \ - \
(\triangle \eta^j ) \, \pa_j \xi^i \ + \ \xi^j \pa_j (\triangle \eta^i) \ - \ \eta^j \pa_j (\triangle \xi^i)
\ + \ 2 \, W_{(\xi,\eta)} \ , \end{eqnarray*}
where we have defined $$ W_{(\xi,\eta)} \ := \ \[ (\^\pa_k \xi^j ) \, \pa_j (\^\pa_k \eta^i) \ + \ (\pa_k \xi^j ) \, \pa_j (\pa_k \eta^i) \ - \ (\^\pa_k \eta^j ) \, \pa_j (\^\pa_k \xi^i) \ - \ (\pa_k \eta^j ) \, \pa_j (\pa_k \xi^i) \] \, \pa_i \ . $$ This computation shows that
\beql{eq:alg4} Z_\vphi \ = \ [ Y_\xi , Z_\eta ] \ - \ [Y_\eta , Z_\xi ] \ + \ 2 \,  W_{(\xi,\eta )} \ . \eeq

Combining this with \eqref{eq:alg3}, the first of \eqref{eq:itosymmshort} reads simply
\beql{eq:alg5} W_{(\xi,\eta)} \ = \ 0 \ . \eeq
But we have already used the condition that $Y_\xi , Y_\eta$ are symmetries of the SDE identified by $X$, hence \eqref{eq:alg5} has no reason to be true in general.

\medskip\noindent
{\bf Example 11.}
In order to check and substantiate this claim, we can consider Example 5 above, i.e. the Ito equation \eqref{eq:example5}; in that case we have seen that symmetry generators are written in the form \eqref{eq:symmexample5}. Let us consider two different symmetries $Y_i$ ($i=1,2$) given by
$$ Y_i \ = \ \[ x \, e^{1/x} \ \b_i (w) \] \, \pa_x \ + \ \[ e^{1/x} \, \b_i (w) \ + \ k_i \] \, \pa_w \ . $$
By explicit computation, we have
$$ [Y_1,Y_2] \ = \ - e^{1/x} \ \( \( k_2 + e^{1/x} \b_2 \) \b_1' \ - \ \( k_1 + e^{1/x} \b_1 \) \b_2' \) \ \( x \,\pa_x \ + \ \pa_w \) \ . $$ This is (in general) \emph{not} in the form \eqref{eq:symmexample5}, and hence it is (in general) \emph{not} a symmetry for the equation \eqref{eq:example5}.

\subsection{Stratonovich equations}

The situation is quite different for equations in Stratonovich
form. This is rather evident comparing \eqref{eq:itosymm} and
\eqref{eq:deteqrndstr} (with our previous computation in
hindsight).

\medskip\noindent
{\bf Lemma.} {\it The Lie-point simple symmetry generators of a
given Stratonovich SDE form a Lie algebra.}

\medskip\noindent
{\bf Proof.} We can proceed as above, and define now the vector
fields \beql{eq:alg11} X_0 \ := \ \pa_t \ + \ b^j \, \pa_j \ , \ \
\^X_k := \^\pa_k \ + \ \s^j_{\ k} \, \pa_j \ . \eeq (Note that
here we change slightly our notation w.r.t. Sect.\ref{sec:strato},
in order to keep uniformity with the previous subsection and to
have a notation better suited to the present task.)

The determining equations \eqref{eq:deteqrndstr} are now written
as \beql{eq:alg12} \[ X_0 , Y_\vphi \] \ = \ 0 \ ; \ \
\[ \^X_k , Y_\vphi \] \ = \ 0 \ . \eeq It is then immediate to
check that if $Y_\vphi$ is given by \eqref{eq:alg2}, and $Y_\xi ,
Y_\eta$ satisfy the determining equations $$ [ X_{0} , Y_\xi
] = 0 = [ X_{0} , Y_\eta ] \ ; \ \ [ \^X_{k} , Y_\xi ]
= 0 = [ \^X_{k} , Y_\eta ] \ , $$  then -- just by Jacobi
identity -- \eqref{eq:alg12} is also satisfied. \EOP

\section{Discussion and conclusions}
\label{sec:conclu}

Symmetry methods are widely recognized as one of our most
effective tools in studying nonlinear deterministic equations
\cite{Olver1,symmref}; the literature devoted to symmetry methods
for stochastic differential equations is comparatively smaller,
and moreover only considers invariance of SDEs (in Ito or
Stratonovich form) only under deterministic transformations.

In this note we have considered -- following the approach by
Arnold and Imkeller in their analysis \cite{LArnold,ArnImk} of
normal forms for SDEs transformations -- the transformations of
SDEs under random diffeomorphsims, i.e. diffeomorphisms depending
on a random (multi-dimensional Wiener) process, and obtained the
determining equations for random Lie-point symmetries of Ito
stochastic differential equations.

The case of Stratonovich equations is also treated, in Section \ref{sec:strato}, and the determining equations are also obtained in this case.

We have also discussed the relation between symmetries of an Ito equation and those of the corresponding Stratonovich one; we have shown that in general -- in particular, at the exception of the case where the matrix $\s (x,t)$ is actually independent of the space variables $x^i$ -- these do \emph{not} admit the same symmetries. The reason for this lies in the actual meaning of the ``correspondence'' between Ito and Stratonovich equations \cite{Stroock}.
On the other hand, an Ito equation and the corresponding Stratonovich one do carry the same statistical information, so that one would expect correspondence between symmetries to hold when considering symmetries of the associated Fokker-Planck equation. This is indeed the case, in a sense made precise by our Proposition in Sect.\ref{sec:correspondence}.

We have considered a number of concrete examples (both in the Ito and the Stratonovich case), choosing equations with a physical significance, and explicitly shown that the determining equations we have written down can be analyzed and explicitly solved, i.e. that our theory is concretely applicable.

As stressed above (see Remark 7), here we only focused on the proper definition of random symmetries of a SDE and on the equations which have to be solved to constructively determine them; that is, we have not considered how the symmetries can be used in the study of the SDE (this appears to be a common feature of a large part of literature devoted to symmetry of SDEs).

On the other hand it seems that the use of symmetries in the
framework of SDEs should go through the same general ideas as in
the case of deterministic equations; that is, beyond any specific
technique, the presence of symmetries suggests first of all that
the analysis will be simpler if using symmetry-adapted
coordinates. A glimpse of this is provided in Example 6 (and
Remark 8) above.

More structured results do exist in the case of deterministic
symmetries of SDEs \cite{Kozlov}; we will investigate in future
work how these result can be extended to the framework of the
random symmetries introduced here.



\end{document}


\title{Erratum for ``Random Lie-point symmetries of stochastic
differential equations'' \\ ({\it J. Math. Phys.} {\bf 58}
(2017), 053503)}

\author{Giuseppe Gaeta\footnote{ORCID: 0000-0003-3310-3455}}
\affiliation{Dipartimento di Matematica, Universit\`a degli Studi
di Milano, via Saldini 50, I-20133 Milano (Italy) \\ {\tt
giuseppe.gaeta@unimi.it} }
\author{Francesco Spadaro\footnote{ORCID: 0000-0002-2313-9131}}
\affiliation{EPFL-SB-MATHAA-CSFT, Batiment MA - Station 8, CH-1015 Lausanne
(Switzerland) \\ {\tt francesco.spadaro@epfl.ch}}

\maketitle

In our recent paper \cite{GS}, due to a regrettable and rather trivial
mistake, a term is missing in the expression (6) for the Ito
Laplacian. The correct formula is, of course
$$ \Delta u \ := \ \sum_{k=1}^n \ \frac{\pa^2 u}{\pa w^k \pa w^k} \ + \
\sum_{j,k=1}^n \ \left( \s \, \s^T \right) \ \frac{\pa^2 u}{\pa
x^j \, \pa x^k} \ + \ 2 \ \sum_{j,k=1}^n \s^{ik} \ \frac{\pa^2
u}{\pa x^j \, \pa w^k} \ . $$ (The reader is alerted that the same
mistake found its way into the recent review paper by one of the
authors \cite{GGPR}.)

This error has no consequence on our general discussion --
conducted in terms of the $\Delta$ operator -- except for Section
VIII (see below); but it does affect the specific computations
occurring in concrete examples and some side remarks.

In particular, the following simple amendments should be inserted in the paper as a consequence to the error in eq.(6):

\begin{enumerate}

\item The final part of {\bf Remark 2} should just read ``does now
also include derivatives w.r.t. the $w^k$ variables, which are of
course absent in (9).''

\item In {\bf Example 1}, the last five lines should read as
follows: ``Plugging this into the first equation, we get $F_t=0$,
hence $F = F(z)$ and any smooth function $\vphi (z)$ of $z = x
-\s_0 t$ provides a simple random symmetry for (34). It should
also be noted that $d z = 0$ on solutions to our equation (34),
see Remark 6.''

\item In {\bf Example 2}, the line after ``we get two equations''
should read as: $$  \psi + z \psi_z = 0 \ , \ \ 2 \psi_t - z
\psi_z = 0 \ . $$
    (The conclusions, i.e. the lines below these equations, are correct.)

\end{enumerate}

Note that Examples 3 \& 4 are unaffected by the error in (6); in
particular, concerning Example 3, any function $\eta (z_1,z_2,t)$
satisfies $\Delta (\eta) = 0$.

Moreover:

\begin{itemize}

\item A misprint was present in the last displayed equation of
Example 3; this should read as follows:
    $$ \frac{\pa\eta_2}{\pa t} \ + \ a_2 \ \frac{\pa \eta_2}{\pa z_2} \ + \
    \frac{a_1}{x_1} \ \frac{\pa \eta_2}{\pa z_1} \ = \ 0 \ ; $$
    this equation admits as solution $\eta_2 (z_1,z_2,t) =
    \xi (z_2 - a_2 t)$,  with $\xi$ an arbitrary function.

\item Corrections should also be introduced in the formulas
relating to Examples 5 \& 6; these would require displaying rather
large formulas and hence we will just alert the reader about this
fact.

\item Examples 7 through 10 are (obviously) unaffected.

\end{itemize}
\bigskip

As mentioned above, the error in (6) has some more substantial
consequence in Section VIII. In fact, the main conclusion reached
there turns out to be wrong: for \emph{simple} (deterministic or
random) \emph{symmetries}, there is a full equivalence between an
Ito and the corresponding Stratonovich equation. In the
deterministic case, this was proved by Unal \cite{Unal}; he also showed
that this is not the case for general symmetries: in particular
for symmetries acting on time as well, there is an auxiliary
condition (amounting to a third order differential equation) to be
satisfied; see Proposition 1 in Unal's paper.

Repeating the computation with the correct form of the Ito
Laplacian (6), one can prove that $\delta^i$ defined in (61) is
identically zero. The full computation will be given elsewhere
\cite{GLjnmp}, but the one for the scalar case is rather simple. In fact, in
this case $ \rho = (1/2) \s_x  \s$. Moreover the second
determining equation (11) guarantees that $\vphi_w = \vphi  \s_x -
\s  \vphi_x$; writing $\vphi_{ww}$ and $\vphi_{wx}$ as
differential consequences of this, and with standard computations,
one easily obtains that
$$\delta \ := \  \vphi \, \rho_x \ - \ \rho \, \vphi_x \ - \ (1/2) \, \Delta \vphi \ = \ 0 \ . $$

Correspondingly, the phrase summarizing the results of Section
VIII in the Conclusions (Section X), i.e. the paragraph starting
with ``We have also discussed the relation...'' (up to ``On the
other hand...'') is also wrong. A correct version of this
statement would read as follows: 

\bigskip\noindent
{\it ``The simple
(deterministic or random) symmetries of an Ito equation and those
of the corresponding Stratonovich one do coincide''}.
\bigskip

We apologize to the readers, and thank the anonymous Referee of \cite{GLjnmp} for pointing out the mistake.
